\documentclass[final,3p,times,twocolumn]{elsarticle}
\usepackage{float}
\usepackage{graphicx}
\journal{Physics Letters A}

\begin{document}
\def\ds{\displaystyle}
\begin{frontmatter}

\title{Discrete Rogue waves in an array of waveguides}
\author{S. Efe, C. Yuce}
\address{ Physics Department, Anadolu University, Eskisehir, Turkey}
\ead{cyuce@anadolu.edu.tr} \fntext[label2]{}
\begin{abstract}
We study discrete rogue waves in an array of nonlinear waveguides. We show that very small degree of disorder due to experimental imperfection has a deep effect on the formation of discrete rogue waves. We predict long-living discrete rogue wave solution of the discrete nonlinear Schrodinger equation.
\end{abstract}


\end{frontmatter}

%% \linenumbers

\section{Introduction}

Rogue waves, sometimes known as freak waves or extreme waves, are waves that appear on a finite background as a result of modulational instability. The height of rogue waves is defined as at least two times higher than the average
surrounding background. Rogue waves were observed long time ago in oceans. The well-known one-dimensional nonlinear Schrodinger equation  (NLS) with
attractive nonlinear interaction is a model equation to investigate rogue waves theoretically. In 1983, Peregrine found an analytic
solution of the nonlinear Schrodinger equation \cite{pereg}. The Peregrine soliton,
limiting case of Kuznetsov and Ma soliton \cite{kuz,ma} and
Akhmediev breather \cite{akhe}, explains how rogue waves appear from nowhere and
disappear without a trace. The Peregrine soliton is formed from slightly modulated uniform background and grows until it reaches its maximum value at a specific time. Then the amplitude of the soliton decreases while the width increases and finally it vanishes. That is why the Peregrine soliton is known as doubly localized wave (localized both in space and time). It is well known that some physical systems such as optics, plasma and ultracold atoms are also described by the nonlinear Schrodinger equation. Therefore, the existence of the Peregrine soliton, or more generally rogue wave, is not restricted to oceans \cite{opt,matter1,opt2}. The experimental realization of Peregrine soliton was first made in an optical system in 2010 \cite{deney} and then in a water wave tank in 2011
\cite{deney2}. These experimental realizations show good agreement between theory and the experiments. The Peregrine soliton is the first order
rational solution of the NLS and the second-order rational solution was studied in \cite{takivv} and observed experimentally in \cite{super}. The ratio of maximum amplitude of the rogue wave to the background amplitude is $3$ for the first order rational solution while it is $5$ for the second order one. Rogue wave solution of the NLS in the presence of disorder is also investigated in \cite{randomNLS}.\\
The extension of rogue waves to discrete systems is possible. Rogue wave
solutions for the discrete NLS, Ablowitz-Ladik equation, Salerno model and Hirota equation have been presented in \cite{AL1,AL2,AL3,AL4,AL5,AL6,AL7,AL8,extreme1,extreme2}. We would like to emphasize that the ratio of maximum amplitude of the wave to the background amplitude exceeds $3$ and $5$ for the first and second order discrete solution of the Ablowitz-Ladik lattice, respectively \cite{AL5}. Bludov, Konotop and Akhmediev considered the discrete nonlinear Schrodinger equation (DNLS) to model an array of nonlinear waveguides \cite{konop}. They constructed a controlled formation of a discrete rogue wave by using a proper choice of initial field amplitude. In a real experiment, very small degree of disorder due to the experimental imperfection always exists. In this paper, we show that it has significant role on the formation and dynamics of discrete rogue waves in an array of nonlinear waveguides. We also predict long-living discrete rogue waves.

\section{Discrete Rogue Waves}

The propagation of an optical field in a tight binding waveguide array can be described by the following discrete nonlinear Schrodinger equation
\begin{equation}\label{temel}
i\frac{d\Psi_j}{dz}=-J_j(\Psi_{j+1}+\Psi_{j-1})+V_j\Psi_j+g|\Psi|_j^{2}\Psi_j
\end{equation}
where $\Psi_j$ is the complex field amplitude at the $j$-th waveguide, $z$ is the propagation direction, $J_j$ is the coupling coefficient between $j$-th waveguide and adjacent waveguides, $\ds{V_j}$ is the propagation constant of the $j$-th waveguide, $g$ is the nonlinear interaction constant. We consider attractive interaction, $g<0$. As noted in \cite{konop}, the solution becomes $(-1)^j\Psi_je^{-4iz}$ if we replace $\ds{g\rightarrow-g}$. There are two conserved quantities in the system. These are total power, $\ds{P=\sum_j|\Psi_j|^2}$, and the total energy. 
Our aim is to find discrete rogue wave solutions of the DNLS. Note that a discrete rogue wave is the strong localization of the total energy over a few lattice sites \cite{konop}.  \\
Suppose the waveguide elements are all identical and waveguide separations are the same in the whole system. Therefore propagation constant $\ds{V_j}$  and the coupling coefficient $J_j$ become site independent. In a real experiment, the distance between waveguide centers are not exactly the same and the waveguide elements are not perfectly identical and very small disorder is therefore always unavoidable. In this case, the deviations of coupling coefficient and propagation constant from their site-independent values are very small. We would like to emphasize that a small degree of disorder as a result of experimental imperfection is always neglected on the theoretical study of waveguides. This is reasonable since almost no contribution comes from such unavoidable disorder on fundamental quantum mechanical effects such as Bloch oscillations, Zener tunneling and Anderson localization. However we will show below that even very weak disorder changes the physics of discrete rogue waves drastically. \\
We consider two different types of disorder: diagonal and off-diagonal disorders. Off-diagonal  disorder is introduced by randomly changing separation between the neighboring sites while the sizes of the waveguides are all identical. As a result of off-diagonal disorder, the coupling coefficient becomes random. Here we assume that the coupling constant without disorder equals to one, $\ds{J_0=1}$. Disorder is introduced by randomly varying the coupling constant such that $J_j\in J_0+\epsilon_1 W$, where $\epsilon_1<<1$ is a small dimensionless constant and $W$ are random numbers with zero-mean distribution in the interval $\ds{\left[-1,1\right]}$. The second type of disorder, i.e. diagonal disorder, arises due to the random choice of the propagation constant. In a real experiment, there always exists very small diagonal disorder due to experimental imperfections. Therefore, the propagation constant $\ds{V_j}$ varies from one waveguide to another according to $\ds{V_j= V_0+\epsilon_2W}$, where $V_0=0$ is set to zero for simplicity, $\epsilon_2<<1$ is a small constant. Since disorder is not deliberately introduced, the constants are assumed to be very small, $\ds{\epsilon_1\sim \epsilon_2 <10^{-2}}$. We remark that both diagonal and off-diagonal disorder are so weak to observe Anderson localization but they play an important role on the formation of the discrete rogue waves. To this end, we note that small next neighboring coupling is always present in a typical experiment. However, we check it has no surprising effect on the formation of discrete rogue waves. So, we neglect higher order couplings in our equation (\ref{temel}). Below we perform numerical computation to study the effect of disorder on the formation and dynamics of discrete rogue waves. Before going further, we define a parameter $\ds{\kappa}$ as the ratio of absolute maximum amplitude the wave can reach to the absolute background amplitude.\\ 
\begin{figure}\label{2}
\includegraphics[width=6.0cm]{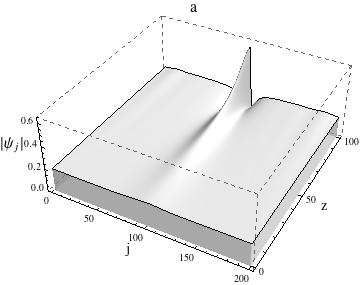}
\includegraphics[width=6.0cm]{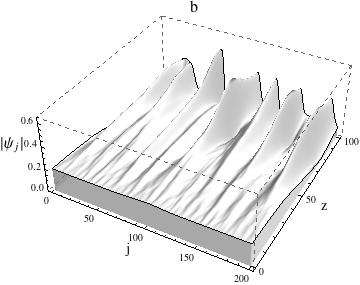}
\includegraphics[width=6.0cm]{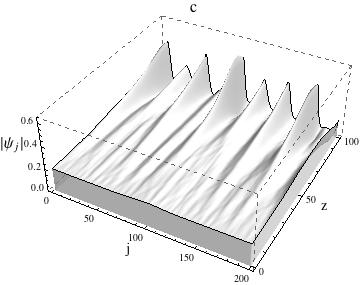}
\caption{The absolute of the field amplitude for $N=200$, $L=100$, $g=-1$ and $A=0.2$.  The initial form of the field is given by (2). The disorder-free case leads to strong localization at the waveguide output as shown in (a). The weak off-diagonal disorder with $\epsilon_1=10^{-2}$ (b) and diagonal disorder with $\epsilon_2=10^{-2}$ (c) due to the experimental imperfection change the spatial evolution drastically. }
\end{figure}
Consider first uniform background density. The system is known as exhibiting modulational instability, which can be investigated by the linearization of the dynamical
equations around the uniform solution, $\ds{\Psi_j=A e^{iqj+iq_z z}}$, where $q$ is the wave number, $A$ is the field amplitude and the nonlinear dispersion relation reads $\ds{q_z=2J\cos(q)-gA^2}$. To examine the stability of this uniform solution, we add a small perturbation by substituting $\ds{\Psi_j\rightarrow \Psi_j+ \delta \Psi_j ~e^{iQj+iQ_z z}}$, where $Q$ is the wave number of the modulation and the dispersion relation $Q_z$ can be found by the linearization of DNLS, $Q_z^2=8 J_0 \sin^2(Q/2) \cos(q)\left( 2J_0 \sin^2(Q/2) \cos(q)+gA^2 \right)$ \cite{offyaa,offyaa2}. We conclude that the plane wave solution is modulationally unstable whenever $\ds{Q_z^2}$ becomes negative. This implies that the unstaggered solution, $q=0$, is unstable if the nonlinear interaction is focusing, $g<0$, while the staggered solution, $q=\pi$, is unstable if the nonlinear interaction is defocusing, $g>0$. We emphasize that the instability is further enhanced in the presence of diagonal and off-diagonal disorder. DNLS equation is non-integrable and the above analysis is therefore approximate. Furthermore, the linearization method shows us the presence of exponentially growing modes while it gives no predictions about the subsequent stages of evolution. To gain more insight, we should perform numerical solution. In the numerical simulations, the system is assumed to be subject to periodic boundary conditions. \\
In \cite{konop}, it was discussed that initial conditions must be chosen properly to observe discrete rogue waves. We first start with a slightly modulated uniform field introduced in \cite{konop}
\begin{equation}\label{ic00}
\Psi_j(z=0)=A\left(1-4\frac{1-2i A^2 L}{1+2A^2 j^2+4A^4L^2}     \right)e^{-iA^2L}
\end{equation}
where $L$ is the length of each waveguide in the array and $A<<1$ is the background amplitude. The Fig-1.a plots the spatial evolution of the above initial wave. The uniform distribution along the lattice is slightly modulated at $z=0$ and then the modulation grows and reaches its maximum at the waveguide output. The total energy is strongly localized into a few waveguides at the waveguide output such that the absolute of the maximum amplitude is $3$ times larger than the background amplitude, $\ds{\kappa=3}$. Therefore the discrete wave can be interpreted as a discrete rogue wave. Let us now investigate the effect of unavoidable weak disorder. Firstly, consider weak off-diagonal disorder with $\ds{\epsilon_1=10^{-2}}$. The presence of weak off-diagonal disorder leads considerable additional energy redistribution among the lattice sites. We see that spatial evolution is drastically changed by small off-diagonal disorder as can be seen from the Fig-1.b. As opposed to disorder-free case, several peaks occur at random locations. Note that the ratio $\ds{\kappa}$ increases in one set of random numbers $W_j$ while it decreases in another set of random numbers $W_j$. To conclude, we say that if we start with the initial field (\ref{ic00}), what we observe in a typical experiment is not like the one in the Fig-1.a but in the Fig-1.b. We numerically check that if we start with the uniform initial wave $\ds{\Psi_j(z=0)=A e^{-iA^2L}}$ instead of (\ref{ic00}), we would observe very similar spatial evolution as in the Fig-1.b. It is generally believed that a proper choice of initial wave, which is generally challenging experimentally, is necessary to observe rogue waves. Here we show that even very weak disorder due to the experimental imperfection plays more dominant role than specific choice of the initial wave on the formation of discrete rogue waves. Let us now consider diagonal disorder. The Fig-1.c plots the spatial evolution of the initial wave (\ref{ic00}) in the presence of diagonal disorder with $\ds{\epsilon_2=10^{-2}}$. As can be seen, very small degree of diagonal disorder changes the spatial evolution similar to the case of off-diagonal disorder. To this end, we say that even tiny amount of disorder with either $\ds{\epsilon_1=2\times10^{-3}}$ or $\ds{\epsilon_2=2\times10^{-3}}$ leads to noticeable effects on the spatial evolution.\\
\begin{figure}[t]\label{20}
\includegraphics[width=6.0cm]{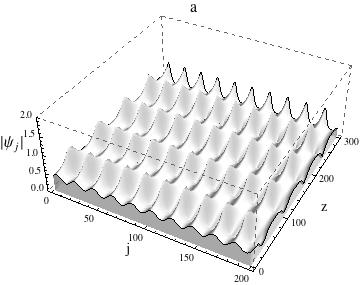}
\includegraphics[width=6.0cm]{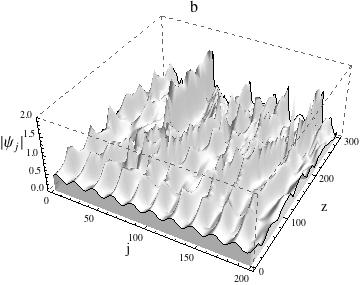}
\includegraphics[width=6.0cm]{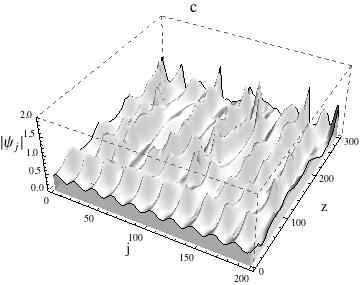}
\caption{ The absolute of the wave function for $N=200$, $\ds{g=-0.5}$, $A=0.4$, $\epsilon=0.2$ and $d=20$. The initial form of the field is given by (3). A doubly periodic discrete rogue wave is formed for the disorder-free case as shown in (a). The doubly periodic character is lost in the presence of weak off-diagonal disorder with $\epsilon_1=10^{-2}$ (b) and diagonal disorder with $\epsilon_2=10^{-2}$. }
\end{figure}
Let us now study another type of discrete rogue waves and analyze the effect of weak disorder. Consider a periodical modulation of uniform field parameter
\begin{equation}\label{ic1}
\Psi_j(z=0)=A\left(1+\epsilon \sin(\frac{2\pi}{d} j)\right)
\end{equation}
where $\ds{A}$ is the background amplitude, $\ds{\epsilon }$ is a small parameter and $\ds{d}$ is the period of the small initial excitation. A question arises. Does modulational instability lead to doubly periodic discrete breather \cite{doubly,doubly2} (discrete rogue waves that is periodic both along the lattice and spatial dimension)? It is not a priori simple problem since the spatial evolution sensitively depends on the parameters. We numerically find that such waves may be formed if $|g|A^2$ is not too big. We see doubly periodical discrete waves with $\ds{\kappa=2.3}$ and $\ds{\kappa=3}$ for the parameters $A=0.4$ and $A=0.5$ when $\epsilon=0.2$ and $g=-0.5$. If we start with the same parameters but $g=-1$, doubly periodic structure would be lost. The Fig-2.a plots doubly periodic discrete wave when $A=0.4$.  The evolution starts with a periodically modulated background and the modulation increases until it reaches its maximum. Then the
system returns to the original background. This repeats periodically in spatial dimension. Let us now study the effect of weak disorder due to the experimental imperfections. The Fig-2.b and Fig-2.c plot spatial evolution in the case of off-diagonal and diagonal disorders, respectively. The modulational instability as a result of small degree of disorder is slower and has almost no effect when $z<50$. Therefore the first cycle remains almost unaffected as shown in the figures 2.b and 2.c. The disorder-related modulational instability dominates at later times and the spatial evolution becomes chaotic. As a result, the wave is not doubly periodic discrete wave anymore but a discrete rogue wave. The presence of disorder increases $\kappa$ from $2.3$ to $3$ with our set of random numbers $W_j$. \\
We have shown that weak disorder plays a vital role on the formation and spatial evolution of discrete rogue waves. Furthermore, an experimental realization of initial waves like (\ref{ic00}) is generally challenging. Therefore, we propose to start with the uniform initial density and let the weak disorder due to the experimental imperfections lead to discrete rogue waves in the system. One can then experimentally observe spatial evolution like the one as in the Fig-1.b and Fig-1.c. Now, a question arises. What is the long time behavior of the discrete rogue waves? One intuitively expects that a couple of peaks appears at random lattice sites and disappears shortly and then another peaks appear in other lattice sites and disappear shortly too. In the same time, chaotic fluctuations happen on the background. As a result, one expects that the positions of the peaks change (not continuously) along the propagation direction. Let us now answer the question. If $z$ is not very large, the dynamics is as what it is expected. Discrete rogue waves with $\kappa \sim3$ appear at random positions and they decay due to the coupling to the neighboring sites. If $z$ is large, there are more peaks in the system and some of them are close enough to each other to start collision. Some collisions in the nonlinear medium excite discrete rogue waves with $\kappa \sim5$. The numerical calculations reveal that the discrete rogue waves with $\kappa \sim5$ don't decay to the adjacent sites. In this way, long-living discrete rogue waves are formed. The Fig-3.a displays long distance spatial evolution of the initially uniform field. As can be seen from the figure, the amplitudes of the three peaks change slightly with $\ds{z}$ while dynamical changes occur chaotically at other lattice sites. The peaks appear from nowhere and almost maintain their profiles. The second message of this paper is the prediction of long-living discrete rogue waves. Let us now understand the origin of such waves. The standard perturbation method fails here since such waves with $\kappa\sim5$ appear at large $z$ and the system is also chaotic. To understand the formation of long-living rogue wave, consider an initial peak at the middle of the lattice with $N=200$. Therefore we suppose the initial field reads $\ds{\Psi_{j}(z=0)=A }$ for $\ds{j\neq100}$ and $\ds{\Psi_{j}(z=0)=pA }$ for $\ds{j=N/2=100}$, where $\ds{p>1}$ is a constant and $A=0.4$. We expect that the excess energy at $\ds{j=100}$ starts to diffract to adjacent sites unless nonlinear interaction is strong enough (if $\ds{gp^2A^2}$ is not larger than $J_0$). If $p$ is large enough (if $p>5$ for our numerical parameters) then nonlinear interaction is dominant and diffraction affects perturbatively. In this case, one can readily say that the amplitude of the peak stays almost the same in the early stage of dynamics. At around $z=50$, a couple of peaks start to appear in the system due to modulational instability. This increases the effect of diffraction on adjacent sites of the peaks. Another question is now arising. Is the peak at $j=100$ destroyed at a large value of $z$? The answer is no as one can see in the Fig.3.b. This can be understood as follows. The coupling of the site $\ds{j=100}$ to the adjacent sites can be enhanced by the existence of another peak at the adjacent sites, which can be formed by modulational instability. It is well known that rogue wave is a rare event and the probability of rogue wave formation exactly at the adjacent sites is very small. Therefore the peak at $\ds{j=100}$ is long-living and can be called a localized discrete soliton on a finite background. Having studied the dynamics of the single initial peak, let us now analyze our original problem. Our original system is initially uniform and small disorder with $\epsilon_2=10^{-2}$ leads to modulational instability. Rogue waves with $\kappa\sim 3$ appear at around $z=100$ as we have already seen in the Fig. 1.c. There are more than one peak in the system and collisions among them lead to higher amplitude peaks. At around $z\sim300$, some peaks with $\kappa\sim5$ appear. As discussed above, they survive very long since their couplings to the adjacent lattice sites are negligible. Additionally, no collisions occurs between these peaks since they don't tunnel to adjacent sites. This prevents the formation of higher order rogue waves with $\kappa>>5$. This is the reason why $\kappa$ starts from one and increases up to nearly $5$ in our numerical simulation. We call it long-living discrete rogue wave since the high-amplitude peaks appear from nowhere and don't disappear.
\begin{figure}\label{off}
\includegraphics[width=6.3cm]{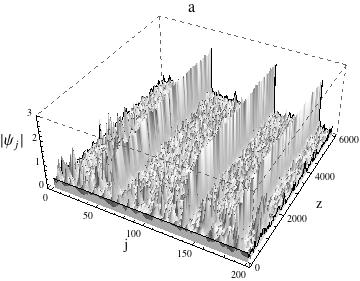}
\includegraphics[width=6.0cm]{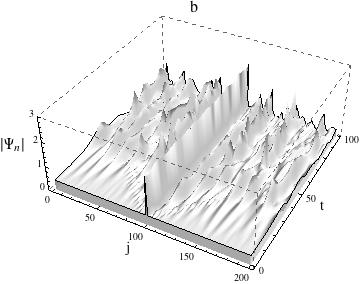}
\caption{The uniform initial field, $\Psi_j(z=0)=A$, leads to long-living discrete rogue waves as can been from the Fig. a. We plot the figure for $0<z<6000$ but the long-living waves survives for a much larger $z$. In the figure (b), a single site initial peak on the uniform background is considered, $\ds{\Psi_{j}(z=0)=A }$ for $\ds{j\neq100}$ and $\ds{\Psi_{j}(z=0)=5A }$ for $\ds{j=100}$. The fig.b show us that if a peak is large enough than the background amplitude, then nonlinear interaction is more dominant than diffraction. Note that the long-living peaks occur due to modulational instability at large $\ds{z}$. The parameters in the figures are given by $\epsilon_2=10^{-2}$, $\ds{g=-1}$, $N=200$ and $A=0.4$. }
\end{figure}

\section{Conclusion}

In this paper, we have studied discrete rogue waves in an array of waveguides. We have discussed that small degree of disorder always exists in a typical experiment. We have considered diagonal and off-diagonal disorders. There are two main results of this paper. Firstly, we find that small degree of disorder leads to significant effects on the formation and dynamics of discrete rogue waves. This is interesting since such a weak disorder has almost no role on Bloch oscillation, Zener tunneling and Anderson localization. Secondly, we predict long-living discrete rogue waves. The long-living discrete rogue wave studied here appears from nowhere and almost maintains its profile instead of vanishing without trace.

\end{document}